\RequirePackage{fix-cm}
\documentclass[smallextended]{svjour3}       

\smartqed 

\usepackage[utf8]{inputenc}
\usepackage{soul}
\usepackage{tikz}
\usepackage{pgfplots}
\DeclareUnicodeCharacter{2212}{−}
\usepgfplotslibrary{groupplots,dateplot}
\usetikzlibrary{patterns,shapes.arrows,shapes.geometric,fit, positioning,arrows.meta,bending}
\pgfplotsset{compat=newest}
\usepackage{multirow}
\usepackage{array}
\usepackage{rotating}
 \usepackage{lineno}
 \usepackage{url}
 \usepackage{tcolorbox}
\newcommand{\realsearchgoal}{\textit{The goal of this research is to aid practitioners in making informed choices about the use of Interactive Application Security Testing (IAST) and Runtime Application Self-Protection (RASP) tools through an analysis of their effectiveness and efficiency in comparison with different vulnerability detection and prevention techniques and tools.}}

\newcommand{\rqIASTeffective}{What is the effectiveness of Interactive Application Security Testing (IAST), as compared to other vulnerability detection techniques, in terms of the number and types of vulnerabilities detected?}

\newcommand{\rqIASTefficiency}{How does the efficiency in terms of vulnerabilities per hour differ between IAST and other vulnerability detection techniques?}

\newcommand{\rqRASP}{What is the effectiveness of RASP in terms of the number and types of vulnerabilities prevented relative to vulnerabilities discovered by vulnerability detection techniques?}

\newcolumntype{C}{>{\centering\arraybackslash}  m}
\newcolumntype{L}{>{\raggedright\arraybackslash}  m}
\newcolumntype{B}{>{\bfseries\raggedright\arraybackslash}  m}
\newcolumntype{T}{>{\centering\arraybackslash\bfseries}  m}
\newcolumntype{O}[2]{%
  >{\begin{turn}{#1}\begin{minipage}{#2}\small\raggedright\hspace{0pt}}c%
  <{\end{minipage}\end{turn}} %
}

\definecolor{cornflowerblue114139207}{RGB}{114,139,207}
\definecolor{lightblue1}{RGB}{132,160,220}
\definecolor{lightblue2}{RGB}{132,160,220}
\definecolor{darkgray176}{RGB}{176,176,176}
\definecolor{lightgray204}{RGB}{204,204,204}
\definecolor{red20400}{RGB}{204,0,0}
\definecolor{red15300}{RGB}{153,0,0}
\definecolor{red234219}{RGB}{234,21,0}
\definecolor{darkred2}{RGB}{126,21,0}

\tikzset{
  comp/.style = {
    minimum width  = 8cm,
    minimum height = 4.5cm,
    text width     = 8cm,
    inner sep      = 0pt,
    text           = green,
    align          = center,
    font           = \Huge,
    transform shape,
    thick
  },
  monitor/.style = {draw = none, xscale = 18/16, yscale = 11/9},
  display/.style = {fill=black!80},
  ut/.style      = {fill = gray}
}
\tikzset{
  computer/.pic = {
    \node(-m) [comp, pic actions, monitor]
      {\phantom{\parbox{\linewidth}{\tikzpictext}}};
    \node[comp, pic actions, display] {\tikzpictext};
    \begin{scope}[x = (-m.east), y = (-m.north)]
      \path[pic actions, draw = none]
        ([yshift=2\pgflinewidth]-0.1,-1) -- (-0.1,-1.3) -- (-1,-1.3) --
        (-1,-2.4) -- (1,-2.4) -- (1,-1.3) -- (0.1,-1.3) --
        ([yshift=2\pgflinewidth]0.1,-1);
      \path[ut]
        (-1,-2.4) rectangle (1,-1.3)
        (-0.9,-1.4) -- (-0.7,-2.3) -- (0.7,-2.3) -- (0.9,-1.4) -- cycle;
      \path[pic actions, fill = none]
        (-1,1) -- (-1,-1) -- (-0.1,-1) -- (-0.1,-1.3) -- (-1,-1.3) --
        (-1,-2.4) coordinate(sw)coordinate[pos=0.5] (-b west) --
        (1,-2.4) -- (1,-1.3) coordinate[pos=0.5] (-b east) --
        (0.1,-1.3) -- (0.1,-1) -- (1,-1) -- (1,1) -- cycle;
      \node(-c) [fit = (sw)(-m.north east), inner sep = 0pt] {};
    \end{scope}
  }
}

\AtBeginDocument{%
  \providecommand\BibTeX{{%
    \normalfont B\kern-0.5em{\scshape i\kern-0.25em b}\kern-0.8em\TeX}}}

\begin{document}



\title{Comparing Effectiveness and Efficiency of Interactive Application
Security Testing (IAST) and Runtime Application Self-Protection (RASP) Tools in a Large Java-based System}
\titlerunning{Comparing  Effectiveness and Efficiency of IAST and RASP}        

\author{Aishwarya Seth        \and
        Saikath Bhattacharya  \and
        Sarah Elder           \and
        Nusrat Zahan          \and
        Laurie Williams
}


\institute{A. Seth \at
              North Carolina State University \\
              \email{aseth@ncsu.edu}           
           \and
           S. Bhattacharya \at
              Milwaukee School of Engineering \\
    \email{bhattacharya@msoe.edu}
           \and
           S. Elder \at
              North Carolina State University \\
            \email{seelder@ncsu.edu}
           \and
            N. Zahan \at
              North Carolina State University \\
              \email{nzahan@ncsu.edu}           
           \and
            L. Williams \at
              North Carolina State University \\
              \email{laurie\_williams@ncsu.edu}           
}



\maketitle
\begin{abstract}
\textbf{CONTEXT:} 
Security resources are scarce, and practitioners need guidance in the effective and efficient usage of techniques and tools available in the cybersecurity industry for detecting and preventing the exploitation of vulnerabilities in software, as per the practitioners' requirements.  Two emerging tool types, Interactive Application Security Testing (IAST) and Runtime Application Self-Protection (RASP), have not been thoroughly evaluated against well-established counterparts such as Dynamic Application Security Testing (DAST) and Static Application Security Testing (SAST). \\ \\
\textbf{OBJECTIVE:}
\realsearchgoal~
\\ \\
\textbf{METHODS:} 
 We apply IAST and RASP on OpenMRS, an open-source Java-based online application. We compare the efficiency and effectiveness of IAST and RASP with techniques applied on OpenMRS in prior work: Systematic (SMPT) and Exploratory (EMPT) Manual Penetration Testing techniques, as well as SAST and DAST tools. We measure efficiency and effectiveness in terms of the number and type of vulnerabilities detected and prevented per hour. \\ \\
\textbf{RESULTS:}
Our study shows IAST performed relatively well compared to other techniques, performing second-best in both efficiency and effectiveness. IAST detected eight Top-10 OWASP security risks compared to nine by SMPT and seven for EMPT, DAST, and SAST. IAST found more vulnerabilities than SMPT. The efficiency of IAST (2.14 VpH) is second to only EMPT (2.22 VpH). These findings imply that our study benefited from using IAST when conducting black-box security testing. We also found RASP only prevents Injection attacks in OpenMRS.
\\ \\
\textbf{CONCLUSION:} 
In the context of a large, enterprise-scale web application such as OpenMRS, RASP does not replace vulnerability detection, while IAST is a powerful tool that complements other techniques. 
\end{abstract}

\section{Introduction}
To build secure software while addressing the ever-growing attack surface, practitioners must utilize the available resources as efficiently as possible to remove the most vulnerabilities from software. Practitioners often use different technologies that optimize resources and increase efficiency to improve vulnerability detection efforts while not expanding the resources. 
Therefore, practitioners can benefit from guidance in selecting vulnerability detection and prevention techniques and tools.

Prior work comparing vulnerability detection techniques, such as the series of studies by Elder et al. and Austin et al. ~\cite{Austin2011,Austin2013,Elder2022}, does not take into account the performance of two emerging tool types, Interactive Application Security Testing (IAST) and Runtime Application Self-Protection (RASP). IAST vulnerability detection tools inject code into, i.e., instrument, the executable form of the application, enabling the tool to scan the source code while \emph{also} collecting dynamic information from real-time interactions with the application~\cite{Pan2019,SASTDASTIAST}.
Similarly, Rajapakse et al. \cite{Rajapakse2021} and Heijstek \cite{heijstek2023bridging} noted that both IAST and RASP are emerging tools for secure DevOps and CI/CD environments and are not well investigated.
On the other hand, RASP tools add another dimension of security to an application in the form of \textit{vulnerability exploitation prevention} by detecting and blocking attacks happening in a real-time production environment.


\realsearchgoal

Ours is one of the initial works to compare IAST and RASP against other automated security testing tools, and we apply these tools to one of the largest production systems. To ensure our results can be compared with prior findings of other techniques, we analyze IAST and RASP tools in the same system, following the methodology specified by Elder et al.~\cite{Elder2022}. The prior work by Elder et al. and Austin et al.~\cite{Austin2011,Austin2013,Elder2022} compared four vulnerability detection techniques - two automated techniques, Dynamic Application Security Testing (DAST) and Static Application Security Testing (SAST), and two manual vulnerability detection techniques- Exploratory Manual Penetration Testing (EMPT) and Systematic Manual Penetration Testing (SMPT). Elder et al.~\cite{Elder2022} replicated the Austin et al.~\cite{Austin2011,Austin2013} work ten years later using these four vulnerability detection techniques on OpenMRS, an open-source medical records system made up of almost four million lines of Java code.

While IAST is a vulnerability detection tool similar to the four techniques examined by Elder et al.~\cite{Elder2022}, RASP focuses on preventing exploitation of vulnerabilities. While vulnerabilities only need to be detected once, preventing exploitation is an ongoing task. Consequently, RASP cannot be reasonably compared to the other techniques in terms of efficiency. For RASP, we focus on the technique's effectiveness in terms of the vulnerabilities it can prevent exploitation of. We compare IAST against vulnerability detection techniques in terms of both effectiveness and efficiency. 


Our work addresses the following \textbf{research questions}:

\begin{description}
  \item[RQ1:] \rqIASTeffective
   \item[RQ2:] \rqIASTefficiency
   \item[RQ3:] \rqRASP
\end{description}

We studied IAST and RASP tools in terms of effectiveness and efficiency and compared our findings with the four techniques and tools used in the Elder et al. study~\cite{SethThesis2022}. We performed our analysis based on the quantity and type of vulnerabilities detected by the IAST tools and the quantity and type of vulnerabilities prevented by the RASP tools to assess the effectiveness of the tools. We compute the efficiency of IAST tools using the quantity of unique and true positive vulnerabilities detected per hour. 

Using IAST, we found 52 vulnerabilities not found in the previous work. The efficiency and effectiveness of IAST were comparable to results from prior work: higher than some techniques and lower than others. RASP, on the other hand, prevented attacks against 44 vulnerabilities but was only effective against Injection attacks.

This work makes the following contributions:
\begin{enumerate}
\item Analysis and comparison of the effectiveness of IAST tools based on quantity and type of true-positive vulnerabilities detected;
\item Analysis and comparison of the efficiency of IAST tools based on the quantity of true positive vulnerabilities \textit{detected} per hour; and
\item Analysis of the effectiveness of RASP tools based on the number of true-positive vulnerabilities whose exploitation was \textit{prevented}.
\end{enumerate}

The remainder of this paper proceeds as follows: Section~\ref{sec:def} provides more detailed explanations of the tools and techniques.
Section~\ref{chap:relwork} describes related work.
Section~\ref{sec:baseline} provides an overview of the work by Elder et al.~\cite{Elder2022}, which we compare our results against. 
Section~\ref{sec:methodology} describes the method employed for answering our research questions.
Section~\ref{sec:results} states and compares the results achieved from our experiments.
Section~\ref{chap:disc} discusses our interpretation of the results obtained, and Section~\ref{sec:lims} presents the limitations of the study.
Section~\ref{chap:conc} concludes, and Section~\ref{sec:future} presents future work.

\section{Technique Definitions}\label{sec:def}
In this section, we provide a more detailed explanation of the vulnerability detection and prevention techniques utilized in our work.
\subsection{Techniques Added in Current Work - IAST and RASP}
In this paper, we focus on two tools: Interactive Application Security Testing (IAST) and Runtime Application Security Protection (RASP). IAST (Section~\ref{sec:desc-IAST}) is used for vulnerability detection, and RASP (Section~\ref{sec:desc-RASP}) is used to actively deter attackers by preventing vulnerabilities from being exploited.

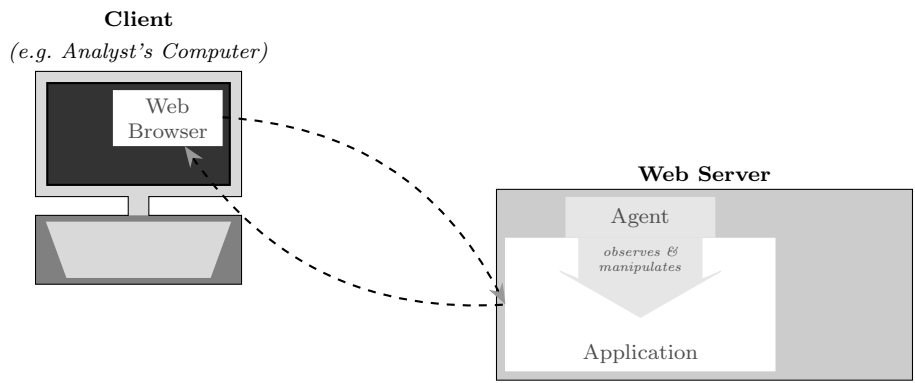
\begin{figure}[htb!]
\centering
\begin{tikzpicture}
\begin{scope}
  \pic(comp0) [
    draw,
    fill = gray!30,
    scale=0.3
  ]
  {computer};
  \node (comp_text1) [anchor=south] at (comp0-c.north) {\textit{(e.g. Analyst's Computer)}};
  \node (comp_text0) [anchor=south] at (comp_text1.north) {\textbf{Client}};

  \node (wb) [rectangle, draw=white, fill=white, color=white, text width=35, anchor=north, text=black!70, font=\footnotesize, align=center, below left=10pt of comp0-m.north east]  {Web Browser};

    \node (s) [rectangle, draw, fill=gray!40, text width=150, minimum height = 72, anchor=north, text=white, font=\footnotesize, text centered, text depth = 1pt,  right=95pt of comp0-c.south east ]  {}; 
    \node (s_text) [anchor=south] at (s.north) {\textbf{Web Server}};

    \node (app) [rectangle, draw, fill=white, text centered, color=white, text width=95, minimum height = 50, anchor=north, text=black!70, font=\footnotesize,  above right =5pt of s.south west]  {};
    \node (app_text) [anchor=south, text=black!70,font=\footnotesize] at (app.south) {{Application}};
    
    \node (agent) [rectangle, draw, fill=white, text centered, color=gray!20, text width=50, minimum height = 15, anchor=south, text=black!70, font=\footnotesize,  above=0pt of app.north]  {Agent};
    \node (aa1) [single arrow, draw,color=gray!20, fill=gray!20, text centered,  text width=40, minimum height = 5,single arrow tip angle=120, anchor=south, text=black!70, font=\tiny,  shape border rotate=270, , single arrow head indent=3pt, below=0pt of agent.south]  {\textit{observes \& manipulates}};



    \draw [draw=gray!80, thick, dashed,
         arrows={Stealth[gray!80,length=8pt,bend,line width=0pt]}-]
    (wb) edge [bend right=30] ($(app.west)$);
    \draw [draw=gray!80, thick, dashed,
         arrows={-Stealth[gray!80,length=8pt,bend,line width=0pt]}]
    (wb) edge [bend left=30] ($(app.west)$);

\end{scope}
\end{tikzpicture}
\caption{Abstraction of IAST and RASP (Java Example)}
\label{fig:Agent}
\end{figure}

Although IAST and RASP achieve different goals, they frequently leverage similar underlying mechanics, shown in Figure~\ref{fig:Agent}. Both tools are integrated into a running application to provide information about system security. The technology used to integrate IAST and RASP tools with the application depends on the technologies and platforms used by the application. For example, in Java systems such as OpenMRS, this integration is typically done through Java Agents. Java Agents are special software that can be used to manipulate the bytecode of a running Java program~\cite{JavaInstrumentationAPI,JavaAgents}. As shown in Figure~\ref{fig:Agent}, interactions with the application by an Analyst, Malicious actor, or other ``Client'' of the web application can be observed or manipulated by the Agent.

\subsubsection{Interactive Application Security Testing (IAST)}\label{sec:desc-IAST}
IAST vulnerability detection tools have access to the code, enabling them to perform static analysis. IAST tools also involve active real-time interaction with the application being tested for vulnerabilities, similar to dynamic analysis. An agent of the tool is integrated with the server of the application so that the tool monitors all the requests, code, and data flow within the application~\cite{Tudela2020}. The code of the application pertaining to those requests and data flow is scanned for vulnerabilities. Therefore, the code coverage provided by an IAST tool is in accordance with how thoroughly the application is being used. Since IAST tools scan code and also use dynamic analysis for detecting vulnerabilities, they are said to incorporate advantages of both SAST and DAST tools.~\cite{Pan2019}

\subsubsection{Runtime Application Security Protection (RASP)}\label{sec:desc-RASP}
RASP tools try to detect and block attacks happening in a real-time production environment. RASP is integrated into the application at the time of deployment. The RASP agent has access to the source code as well as the ability to control the execution of the application~\cite{Fry2021}.  When a malicious attempt is made to change the behaviour or state of the application, the RASP agent actively tries to detect and block the attack. Thus, RASP tools add another dimension of security to applications in the form of \textit{vulnerability exploitation prevention}, unlike other tools discussed in this paper which are used in the testing phase for \textit{vulnerability detection}. 

RASP should not be confused with Web Application Firewall (WAF). 
RASP monitors input, output, and data flow to detect an attack and proceeds to block the attack detected.
On the the other hand, WAF monitors input requests only and does naive pattern matching for detection of malicious activity. They do not have access to the source code of the application and require extensive configuration testing to cover an application sufficiently.~\cite{Cisar2016}



\subsection{Techniques from Prior Work - SMPT, EMPT, SAST, and DAST}
We compare our results against those of prior work~\cite{Elder2022}, which has examined four vulnerability detection techniques. 

\begin{description}
    
    \item[\textbf{Systematic Manual Penetration Testing (SMPT)}]: A type of testing technique that ISO 29119-1 defines as “dynamic testing in which the [analyst]’s actions are prescribed by written instructions in a test case" ~\cite{2013ISO29119-1}.
    
    \item[\textbf{Exploratory Manual Penetration Testing (EMPT)}]: A type of testing technique in which the analyst “spontaneously designs and executes tests based on the [analyst]’s existing relevant knowledge”~\cite{2013ISO29119-1}, searching for vulnerabilities.
    
    \item[\textbf{Static Application Security Testing (SAST)}]: A type of automated testing in which the source code of the application is analyzed to find security flaws in the application~\cite{mcgraw2006software}. 
    
    \item[\textbf{Dynamic Application  Security Testing (DAST)}]: A type of automated testing in which dynamic analysis of the application is used to find vulnerabilities. DAST involves no access or knowledge of the application's source code~\cite{Elder2022}.

\end{description}

\section{Related Work}~\label{chap:relwork}
Our study focuses on comparing the effectiveness and efficiency of IAST vulnerability detection tools with SMPT, EMPT, SAST, and DAST; and determining the relative effectiveness of RASP. There is related work in the evaluation of vulnerability detection tools and techniques, which we discuss in this Section. However, few studies have examined IAST and RASP, and no studies have examined the full breadth of these techniques in a large-scale production system.   

\subsection{Comparing Static Application Security Testing (SAST) and Dynamic Application Security Testing (DAST)}

SAST and DAST tools have been benchmarked and compared (e.g., \cite{Antunes2009,antunes2010,antunes2014,antunes2015}) based upon their effectiveness in the web services domain. These studies found that SAST is more effective than DAST in finding vulnerabilities. However, their bench-marking approach requires manual code review, which does not scale.  Other studies, such as \cite{Higuera2020}, focus on SAST tools for detecting vulnerabilities and find that the true positive and false positive rates vary widely across tools and across different types of vulnerabilities. Croft et al.~\cite{croft2021} compares open-source, rule-based SAST tools with learning-based software vulnerability prediction models for C/C++ software systems. Croft et al.\cite{croft2021} concluded that although learning-based approaches had better precision, both learning-based and SAST tools approaches should be used independently. Piskachev et al. \cite{piskachev2023can}  did a user study of SAST tools in resolving security vulnerabilities and provided a list of recommendations for software security professionals and practitioners.

Scandariato et al.~\cite{Scandariato2013} studied users' experiences of nine participants using a SAST tool and an automated tool for penetration testing on two blogging applications. The participants stated that they found SAST to be more effective for security teams. 


Fonseca et al.~\cite{Foncesca2007} performed an empirical evaluation of the effectiveness of different DAST tools for finding SQL and XSS vulnerabilities in web applications. They found that almost all the tools suffered from high false-positive rates and low coverage.


In a comparison study of eleven DAST tools by Doupe et al.~\cite{Doupe2010} on a web application, the researchers found that some vulnerabilities can be detected reliably, whereas others were more difficult for the tools to detect. However, their study was more focused on whether the tools could reach `deep' parts of code via crawling and detect logic vulnerabilities. 

 Bau et al.~\cite{Bau2013} have compared four DAST tools based on vulnerabilities detected per thousand lines of code on 27 web applications. They conclude that more than one tool should be used and that the majority of the 'Information Leak' alerts were false positive. Amankwah et al.~\cite{Amankwah2020} compared eight commercial and open-source web DAST tools on two web applications. They compare the scanners based on precision, recall, Youden index, OWASP web benchmark evaluation (WBE), and the web application security scanner evaluation criteria (WASSEC). They find that although both open-source and commercial DAST tools can be effective, no single tool was effective in finding all the vulnerabilities.

 Cruz et al.~\cite{cruz2023open} compare Open-Source SAST, DAST, and Software Composition Analysis (SCA) tools. The authors found combinations of tools were more effective than using one particular tool or approach. While the authors mention IAST, including the fact that one of their tools included IAST components, as with other comparisons - they focus their evaluation on SAST, DAST, and SCA without evaluating IAST.

Rajapakse et al. \cite{Rajapakse2021} investigated issues during the integration of security tools into a DevOps workflow by software practitioners. Heijstek \cite{heijstek2023bridging} also mentioned that both IAST and RASP are emerging tools for secure DevOps and CI/CD environments.  Tudela et al. \cite{MateoTudela2020OnCS}  worked on combining SAST, DAST, and IAST security analysis techniques, applied against the OWASP Benchmark project.   Miao et al. \cite{Miao2023} provided stratigies of integrating RASP  protection policies in a security information and event management framework. 

 

\subsection{Interactive Application Security Testing (IAST)}

Interactive Application Security Testing (IAST) is an automated vulnerability detection tool. Setiawan et al. ~\cite{Setiawan2020} implemented an IAST approach and tested their approach on a government website and found 249 vulnerabilities covering all the categories of the OWASP Top Ten (2017). However, their approach does not accurately represent the approaches 
employed by commercially available IAST tools for vulnerability detection. Their approach uses a SAST tool that scans the entire code of the application while a DAST tool simultaneously does dynamic analysis of the application. In contrast, modern IAST tools do not scan the entire source code of an application. They only scan the parts of the application's source code that interact with users through test cases. In this study, we study two commercial IAST tools and compare their performance with other vulnerability detection techniques and tools.
Tudela et al.~\cite{Tudela2020} compared the effectiveness of SAST, DAST, and IAST tools by running them against the OWASP Benchmark project. The OWASP Benchmark project~\cite{OWASPBenchmark} is a collection of thousands of test cases developed primarily in Java. They chose two tools of each SAST, DAST, and IAST and evaluated the effectiveness of each tool separately, in pairs, and finally in groups of three. They found that the IAST-IAST pair yields better results and that combining IAST with either DAST or SAST improves vulnerability detection effectiveness. The study was primarily focused on how tools may be paired to increase effectiveness. However, in this work, we studied a large-scale application  (OpenMRS), which employs a wide range of technologies and has millions of lines of code, and individually compared the effectiveness of the techniques and tools.

\subsection{Runtime Application Self Protection (RASP)}
Runtime Application Self Protection (RASP) is emerging tool for preventing the exploitation of vulnerabilities. 
Most of the studies on RASP have been related to runtime protection mechanisms in specific types of applications, with RASP tools implemented by researchers themselves. In contrast, our current analysis focuses on RASP tools used in industry. 

Pupo et al.~\cite{Pupo2021} developed a two-phase abstract interpretation approach, extracting SAST components from two RASPs. They evaluate their approach by comparing their two-phase approach with a single-phase RASP approach and find that the two-phase approach is better than the single-phase approach in terms of precision. Similarly, Huang et al.~\cite{Huang2004} developed a framework called WebSSARI for PHP web applications that incorporates static analysis for finding potentially insecure code and runtime protection by inserting runtime guards in potentially insecure code. They test their framework on 11,848 files from 230 open-source projects to find and prevent exploitation of vulnerabilities. 

Yin et al.~\cite{Yin2018} developed a RASP tool specific to Script Injection attacks for dynamic web applications based on data flow analysis and automatic insertion of filters before relevant sink statements. They compare their solution with WAF, BEEP, and CSP and find that their approach and CSP perform better. However, CSP was expensive. 

Yuan et al. and Bailey et al.~\cite{Yuan2013,Bailey2014} described a self-adaptive framework for RASP which provides defense-in-depth detection and takes action to block the attack from requirements-driven or architecture-based approaches. 

\section{Baseline comparison:  Elder et al.~\cite{Elder2022}}
\label{sec:baseline}
We directly compare our results with the techniques from the previous study by Elder et al.~\cite{Elder2022}. The four techniques and tools examined by Elder et al.~\cite{Elder2022} are SMPT, EMPT, SAST, and DAST. Section \ref{sec:elderOverview} provides information from their methodology, which guided our own methodology as described in Section~\ref{sec:methodology}.  Section \ref{sec:elderFindings} briefly summarizes Elder et al.'s results. 

\subsection{Baseline Research Methodology}
\label{sec:elderOverview}
 Elder et al.~\cite{Elder2022} replicated prior work by Austin et al.~\cite{Austin2011,Austin2013}  comparing the SMPT, EMPT, SAST, and DAST vulnerability detection techniques and tools.  Elder et al.~\cite{Elder2022} applied each technique to the Open Medical Records System (OpenMRS) web application, an open-source, Java-based electronic medical records management system of about 4 million lines of code. They compare the techniques and tools based on their effectiveness and efficiency. They measure the effectiveness of a vulnerability detection technique/tool based on the number and type of vulnerabilities detected by that technique/tool. Efficiency is measured in terms of the number of vulnerabilities detected per hour.  We provide high-level information on their research approach that is specific to the four vulnerability detection techniques.  The vulnerability de-duplication, true/false positive classification, and vulnerability counting processes used are the same as those outlined in Section \ref{sec:procedure}.
\begin{description}
    \item[SMPT:] 131 black box security test cases were developed and executed.  These test cases were based upon the OWASP Application Security Standard (ASVS)\footnote{https://owasp.org/www-project-application-security-verification-standard/}.   Two independent analysts executed each test case. 
    \item[EMPT:] Sixty-three (63) students in a graduate-level software security class each spent three hours performing exploratory penetration testing as part of their final project submission at the end of the semester. 
    \item[SAST:]  One open source (Sonarqube version 8.2) and one proprietary SAST tool were run on OpenMRS using the default security rules. Two analysts independently reviewed the alerts that the tools produced as output.
    \item[DAST:] One open source (OWASP ZAP) and one proprietary DAST tool were run. DAST requires sample inputs to the application. In Elder et al., the sample inputs to the DAST application were based on six test cases from the SMPT test suite. The test cases were selected to maximize the coverage of testing of OpenMRS.  Two analysts independently reviewed the alerts that the tools produced as output. 
\end{description}

The results of running these four vulnerability detection tools were recorded in a dataset containing the following details for each vulnerability recorded:

\begin{enumerate}
    \item Request URL and parameters.
    \item Source Code Location: Consisted of module name, file name, and line number.
    \item CWE number.
    \item OWASP Top Ten (2021) category.
    \item The vulnerability detection technique or tool which detected it.
    \item Description: Additional descriptive information required to distinguish the vulnerability from other vulnerabilities
\end{enumerate}

For measuring efficiency, Elder et al. used data collected from a graduate-level software engineering course to reduce potential biases introduced by collecting efficiency data from a single source. For tool-based techniques, DAST and SAST, students were instructed to record the time required to review a subset of the tool results. For SMPT, students wrote and executed a series of test cases over several assignments, recording the amount of time required and the number of vulnerabilities found. For EMPT, students performed exploratory testing for 3 hours each, recording the number of vulnerabilities found.  In the current study, we compare our results against the average VpH recorded for each technique. Details on the assignments, how data was collected, and additional analysis of the results can be found in the original study~\cite{Elder2022}.

\subsection{Results}
\label{sec:elderFindings}
Elder et al.~\cite{Elder2022} found the most vulnerabilities using SAST tools but found more severe vulnerabilities using EMPT. They discovered vulnerabilities using each technique that were not found by the others. In terms of efficiency, they found that the efficiency of manual techniques, that is, EMPT and SMPT, was comparable to or better than the efficiency of automated techniques, that is, DAST and SAST, in terms of Vulnerabilities per Hour (VpH). Results for each of the four vulnerability detection techniques will be presented in Section~\ref{sec:results}, where we compare our IAST and RASP results with those of Elder et al.~\cite{Elder2022}.

\section{Methodology}
\label{sec:methodology}
Our research questions require measurement of the efficiency and effectiveness of IAST and RASP tools comparable to prior work with SAST, DAST, SMPT, and EMPT. Specifically, our first research question, on the effectiveness of IAST tools, states: \textit{RQ1: \rqIASTeffective }
We define \textit{effectiveness} to include (1) the number of true positive vulnerabilities detected; and (2) the spectrum of Common Weakness Enumeration (CWE) types and OWASP Top Ten types for the vulnerabilities detected. 

Our second research question on the efficiency of IAST tools states that:
\textit{RQ2:  \rqIASTefficiency}
Consequently, we determine the number of ‘True Positive’ and ‘Unique’ vulnerabilities detected per hour to measure the efficiency of IAST tools.

Our third research question focuses on the effectiveness of the RASP technique and states that:
\textit{RQ3: \rqRASP } 
Similar to our RQ1, we define the \textit{effectiveness} of RASP tools to include (1) the number of true positive vulnerabilities prevented, which were detected by the vulnerability detection techniques and tools; and (2) the spectrum of Common Weakness Enumeration (CWE) types and OWASP Top Ten types.


\subsection{Data Collection}
\label{sec:data}

Since the working of IAST and RASP tools is such that they require interaction with the application, we use the same 131 SMPT black-box test cases as those developed and used by Elder et al.~\cite{Elder2022} to systematize the interactions with our application. The SMPT test cases were based on ASVS Level Control 1 since it is the only level that is ``completely penetration testable by humans''~\cite{Elder2022}. The students and researchers participating in their study compiled a set of 131 test cases covering 63 controls of ASVS.

We use two IAST tools, namely, IAST-1 and IAST-2, and one RASP tool. We perform the interactions on the three tools running on the OpenMRS application, hosted on a virtual image. We use OpenMRS version 2.9 to enable comparison with Elder et al.~\cite{Elder2022}. We do a detailed comparison of each individual vulnerability detected by IAST and RASP to those found in their study for every technique and tool. For the classification of alerts, we refer to the same version of the source code of the application as Elder et al.~\cite{Elder2022}.


\subsection{Vulnerability Counting Guidelines}
\label{sec:count}
For both IAST and RASP, multiple alerts may point to the same vulnerability. We use the definition of vulnerability specified by NVD and Common Vulnerability Enumeration (CVE), which is the source of the vulnerability database of NVD \cite{NVDVulnDef}. Therefore, we use the CVE Counting Rules~\cite{MITRECVECount}, the same guidelines as CVE Numbering Authorities (CNA), to classify and count the `true/false positive' and `unique' alerts. Key guidance in determining which alerts represented unique vulnerabilities is as follows:

\begin{itemize}
\item Each unique vulnerability must be independently fixable
\item If there are multiple alerts of the same vulnerability type, but none of them share vulnerable code (i.e., they must be fixed independently), then all of them are considered `unique'.
\item In case of ambiguity regarding the independent fixability of two alerts, only one is considered to be `unique' and the other is considered a `duplicate'.
\end{itemize}

We refer to the protection mechanisms or mitigation suggested by the tools themselves as well as those specified in the OWASP Top Ten cheat sheet~\cite{OWASPCheatSheet} to help understand how a vulnerability would be fixed and whether vulnerabilities are independently fixable.

\subsection{Procedure for Generating and Curating set of Vulnerabilities for IAST}
\label{sec:procedure}

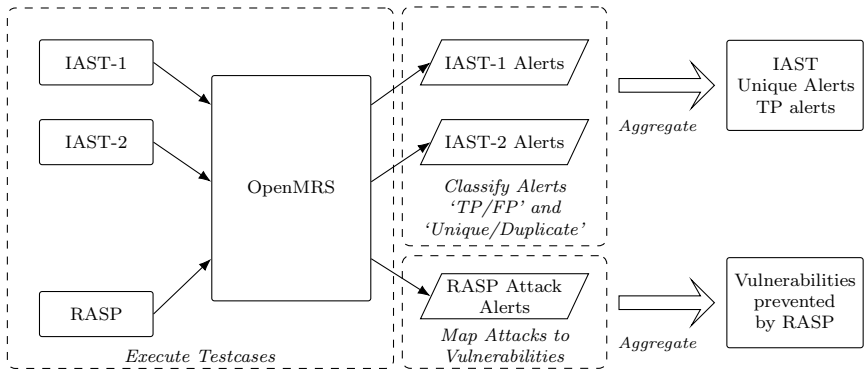
\begin{figure}[htb!]
\centering
\begin{tikzpicture}[scale=0.85, transform shape]
\begin{scope}[auto,
stage_node/.style={rectangle, dashed, rounded corners=3pt, align=center, inner sep=0pt},
system_node/.style={rectangle, rounded corners=1pt, align=center, inner sep=0pt},
result_node/.style={trapezium, trapezium left angle=65, trapezium right angle=115},
arrow_node/.style={single arrow,single arrow head indent=4pt, minimum height = 42pt,minimum width=2pt}
]
\node[draw,stage_node,text width=170pt, minimum height = 160pt, draw](tc) {};
\node (tc_header) [anchor=south] at (tc.south) {\textit{Execute Testcases}};

\node[draw,system_node,text width=50pt, minimum height = 20pt, below right=20pt of tc.north west](i1) {};
\node (i1_text) [anchor=center] at (i1.center) {IAST-1};

\node[draw,system_node,text width=50pt, minimum height = 20pt, below=15pt of i1.south](i2) {};
\node (i2_text) [anchor=center] at (i2.center) {IAST-2};

\node[draw,system_node,text width=50pt, minimum height = 20pt, above right=20pt of tc.south west](r0) {};
\node (r0_text) [anchor=center] at (r0.center) {RASP};

\node[draw,system_node,text width=70pt, minimum height = 100pt, left=10pt of tc.east](om) {};
\node (om_text) [anchor=center] at (om.center) {OpenMRS};


\node[draw,result_node,text width=10pt, minimum height = 20pt, right=122pt of i1.east ](ir1) {}; 
\node (ir1_text) [anchor=center] at (ir1.center) {IAST-1 Alerts};

\node[draw,result_node,text width=10pt, minimum height = 20pt, right=122pt of i2.east ](ir2) {};
\node (ir2_text) [anchor=center] at (ir2.center) {IAST-2 Alerts};


\node[draw,stage_node,text width=90pt, minimum height = 50pt, below=40pt of ir2.south](mv) {};
\node (mv_header) [anchor=south,text width=72pt, align=center] at (mv.south) {\textit{Map Attacks to Vulnerabilities}};

\node[draw,result_node,text width=10pt, minimum height = 20pt, below=8pt of mv.north ](ra) {};
\node (ra_text) [anchor=center,text width=52pt,align=center] at (ra.center) {RASP~Attack Alerts};

\node[draw,stage_node,text width=90pt, minimum height = 106pt,anchor=center, above=4pt of mv.north](ca) {};
\node (ca_header) [anchor=south,text width=82pt, align=center] at (ca.south) {\textit{Classify Alerts `TP/FP' and `Unique/Duplicate'}};
\node[draw,arrow_node, right=40pt of ir1.south east ](a1) {};
\node (a1_text) [anchor=west,font=\scriptsize, above left=-30pt of a1.south west] {\textit{Aggregate}};

\node[draw,arrow_node, below=90pt of a1.south ](a2) {};
\node (a2_text) [anchor=west,font=\scriptsize, above left=-30pt of a2.south west] {\textit{Aggregate}};

\node[draw,system_node,text width=60pt, minimum height = 40pt, right=5pt of a1.east](ia) {};
\node (ia_text) [anchor=center,text width=55pt, align=center,right=0pt of ia.west] {IAST Unique~Alerts TP~alerts};

\node[draw,system_node,text width=60pt, minimum height = 40pt, right=5pt of a2.east](vp) {};
\node (vptext) [anchor=center,text width=55pt, align=center] at (vp.center) {Vulnerabilities prevented by RASP};

    \draw [draw,-{Latex}]
    ($(i1.east)$) -- ($(om.west)+(0,1.3)$);
    \draw [draw,-{Latex}]
    ($(i2.east)$) -- ($(om.west)+(0,0.1)$);
    \draw [draw,-{Latex}]
    ($(r0.east)$) -- ($(om.west)+(0,-1.1)$);
    \draw [draw,-{Latex}]
    ($(om.east)+(0,1.3)$) -- ($(ir1.west)$);
    \draw [draw,-{Latex}]
    ($(om.east)+(0,0.1)$) -- ($(ir2.west)$);
    \draw [draw,-{Latex}]
    ($(om.east)+(0,-1.1)$) -- ($(ra.west)$);


\end{scope}
\end{tikzpicture}
\caption{A high-level overview of our approach.}
\label{fig:approach}
\end{figure}

Figure~\ref{fig:approach} provides a high-level overview of our methodology described in this section, as well as the methodology for RASP described in Section~\ref{sec:raspMeth}.

As part of answering RQ1 and RQ2, two researchers independently executed the following set of steps for both IAST tools:
\begin{enumerate}
    \item Run the 131 black box SMPT test cases used in the previous study by Elder et al.~\cite{Elder2022}
    \item Record how much time is spent on Steps 3-4
    \item \textit{De-duplication.} As detailed in Section~\ref{sec:count}, the tools provide the CWE for each alert. Then, We leverage the information provided by the tool, such as protection mechanisms or other suggested mitigations, as well as information specified in the OWASP Top Ten cheat sheet~\cite{OWASPCheatSheet} for finding a possible fix for a vulnerability. For each alert, we determine if the alert either maps to a different CWE or if the CWE is the same, then determine if a fix for the alert is different from a fix of an already-examined alert. If either of the two conditions is met, we mark the alert as `unique’; otherwise, we mark it as a `duplicate’.
    \item \textit{True positive identification.} As detailed in Section~\ref{sec:tpfp}, we go through each alert to determine if we could exploit the tool alert based on our experience and simulate it in the OpenMRS application OR if the tool alert did not have any protection mechanism in place in the application to prevent exploitation. 
    If either of the two conditions is met, we mark the alert as a `true positive' or otherwise as a `false positive'.
    \item We leveraged the findings from the Elder et al. study for the ``true positive'' AND ``unique'' alerts (from Steps 3 and 4) and thoroughly compared vulnerability to vulnerability from our result set to their result set to identify which vulnerabilities were found using a different technique in their research. For each vulnerability in our set, we examine whether there is a vulnerability in the result set from Elder et al.~\cite{Elder2022} with the same URL, parameter, and CWE type. If the fields match, we mark our vulnerability as the same vulnerability as the one found by Elder et al. and record the technique or tool that detected that vulnerability from the Elder et al.~\cite{Elder2022} study.
\end{enumerate}

Upon completion of these steps for each tool, each independent researcher aggregated the set of `true positive' and `unique' alerts generated from the IAST-1 and IAST-2 tools and removed the common alerts to have a final set of `true positive' and `unique' alerts representing the vulnerabilities detected by IAST technique.

Since the classification of alerts as `true/false positive' or `unique/duplicate' is subject to domain knowledge, experience, and human error, two researchers independently execute the above procedure for classifying the alerts and reviewing each other's results. A third reviewer resolves any conflicts regarding the classification.

\subsubsection{True/False Positive Classification}
\label{sec:tpfp}

We classify an alert generated by the tools as `true positive' when we perceive it to be an actual vulnerability. The National Vulnerability Database (NVD) ~\cite{NVDVulnDef} defines a vulnerability as: ``A weakness in the computational logic (e.g., code) found in software and hardware components that, when exploited, results in a negative impact to confidentiality, integrity, or availability''. Thus, we consider an alert to be `true positive' if the weakness noted by the alert can be exploited such that it can lead to a security breach of the application. However, we take an additional aspect of `Defense in Depth' into consideration to classify the alerts, which highlights the importance of placing different safeguards at multiple layers for better protection. Thus, we also consider if the weakness noted by the alert has any protection mechanism in place in the source code of the application. If no countermeasures have been taken to prevent exploitation of the weakness highlighted by the alert, then we mark the alert as `true positive' even if we are unable to determine whether the weakness can be exploited to breach the security of the application. For example, if a test case generates an XSS alert and we see no input validation is performed in the source code or if we are able to input a JavaScript string that successfully gets injected into and displayed on a page of the application, then we mark the alert as `true positive'.

\subsubsection{Vulnerability Type}
\label{sec:vtype}

We assign the vulnerability types to each `true positive' alert based on two categories - CWE and OWASP Top Ten. “CWE is a community-developed list of software and hardware weakness types.”~\cite{MITRECWE2022}. The OWASP Top Ten is a regularly updated list of “the most critical security risks to web applications.”~\cite{OWASP2022}. We use the OWASP Top Ten 2021 to summarize the CWE types of vulnerabilities found. Each of these Top 10 is comprised of a set of CWE types. The latest (2021) Top Ten, which were used in our analysis, are A01 - Broken Access Control, A02 - Cryptographic Failures, A03 - Injection, A04 - Insecure Design, A05 - Security Misconfiguration, A06 - Vulnerable and Outdated Components, A07 - Identification and Authentication Failures, A08 - Software and Data Integrity Failures, A09 - Security Logging and Monitoring Failures, and A10 - Server-Side Request Forgery (SSRF).   

The tools assign a CWE type to each alert. In cases where two tools assign different CWE types to the same vulnerability, the CWE type is analyzed, and a manual revision of the CWE type may be done. For example, a test case generated a clickjacking alert with CWE type listed as CWE-451 in IAST-1, while the same alert was generated with CWE type listed as CWE-693 in IAST-2. We resolve this conflict by manually updating the value of the CWE type in the final set of vulnerabilities, depending on the test case. 

OWASP Top Ten (2021) Categories were already assigned to each alert by the tools. If an alert had no OWASP Top Ten (2021) category assigned by the tool, we manually assigned the values by mapping from the CWE types. Since OWASP Top Ten focuses on the top ten categories of vulnerability types that are most frequently found in software, it helps in presenting a more relevant and understandable way to interpret the vulnerabilities found.

\subsubsection{Vulnerability Severity}
\label{sec:vseverity}
We determine the importance and severity of vulnerabilities found in two different ways. First, the tools classify the alerts based on their severity. Since each tool has its own classification rule and severity categories, we organize the vulnerabilities into two groups, `more severe' and `less severe' to make our severity categories more consistent. We use the value of severity initially assigned to the set of vulnerabilities found by a tool. The vulnerabilities that are classified as `Low' or below by the tool are grouped as `less severe', whereas those vulnerabilities that are classified by the tool in categories above `Low' are grouped as `more severe' vulnerabilities. Second, since OWASP Top Ten ranks vulnerabilities based on their criticality as well as frequency, we use the OWASP Top Ten categories to understand the importance of the vulnerabilities found by each tool.

\subsubsection{Efficiency}
\label{sec:efficiencyMeth}
As noted in Step 2 above, we record the amount of time spent by each researcher to analyze the alerts produced by each tool and determine the number of true positive vulnerabilities found, removing false positives. For each researcher, we divide the number of true positive vulnerabilities found by the length of time recorded to determine the Vulnerabilities per Hour (VpH) of IAST. We report the average VpH in our results.

\subsection{Procedure for Curating RASP tool Alerts}
\label{sec:raspMeth}

Since RASP is a real-time attack prevention tool, we employ a slightly different methodology than the one adopted for IAST (in our study) and other vulnerability detection techniques and tools (by Elder et al. \cite{Elder2022}). We want to find out the number of vulnerabilities previously detected by the vulnerability detection techniques and tools, but RASP has prevented them from being exploited. We run the same set of 131 black box SMPT test cases as used by Elder et al.~\cite{Elder2022} and in our IAST study. The alerts generated by the RASP tool are for the attacks detected and blocked by the RASP agent. We measure the effectiveness of the RASP tool by the following two parameters: (1) the number of unique vulnerabilities prevented from being exploited by the RASP tool and (2) the type of that vulnerability.

We use the type of attacks blocked by the RASP tool as an initial value to determine the type of vulnerability that was prevented from being exploited. We further map the types of vulnerabilities based on CWE to OWASP Top Ten (2021) categories. The mapping of CWE to OWASP Top Ten helps in determining the number of the most critical vulnerabilities that are being prevented from being exploited by the RASP tool.

To count the number of vulnerabilities prevented by the RASP tool relative to the vulnerabilities detected by the vulnerability detection techniques and tools, we map each of the attack alerts generated by the RASP tool to the vulnerabilities detected by each of the vulnerability detection techniques and tools. An overview of the approach for RASP is depicted in Figure ~\ref{fig:approach}.

\section{Results}
\label{sec:results}

In this section, we outline the results we obtained via the experiments performed using the IAST and RASP tools. Table~\ref{tab:summary} provides a high-level summary of our results for RQ1 - \textit{\rqIASTeffective}, RQ2 - \textit{\rqIASTefficiency} and RQ3 - \textit{\rqRASP}. Detailed results for RQ1, RQ2 and RQ3 are provided in Sections~\ref{sec:results-rq1}, \ref{sec:results-rq1}, and \ref{sec:results-rq1} respectively. We will discuss the implications for our results in the next section (Section~\ref{chap:disc})


\newcommand*{\techColwidth}{0.08\textwidth}
\newcommand*{\curColwidth}{0.1\textwidth}
\begin{table}[ht!]
\centering
\setlength\tabcolsep{1pt}
\renewcommand{\arraystretch}{1.2}
\begin{tabular}{|B{0.36\textwidth}||T{\curColwidth}||T{\curColwidth} | | C{\techColwidth}|C{\techColwidth}|C{\techColwidth}|C{\techColwidth}|}
\cline{2-7}
 \multicolumn{1}{c|}{} & \multicolumn{2}{c||}{\textbf{Current Study}} & \multicolumn{4}{c|}{Previous Study}\\ \cline{2-7}
\multicolumn{1}{c|}{} &  IAST & RASP & SMPT & EMPT & SAST & DAST \\\hline
\hline
Effectiveness:{\newline}\# Vulnerabilities & 91 & 44 & 37 & 185 & 23 & 823\\\hline
Effectiveness:{\newline} \# OWASP Top 10 Covered & 8 & 1 & 9 & 7 & 7 & 7 \\\hline
Efficiency:{\newline}Average VpH & 2.14 & N/A & 0.69 & 2.22 & 0.55 & 1.17\\\hline
\end{tabular}
\vspace{1pt}
\caption{Effectiveness and Efficiency of IAST and RASP compared to other techniques (\textit{Section~\ref{sec:results}})}
\label{tab:summary}
\end{table}

\subsection{RQ1: Effectiveness of IAST}\label{sec:results-rq1}

In this section, we outline the results for our first research question, which is based on the effectiveness of IAST tools: ~\textit{RQ1: \rqIASTeffective}. We provide the number of `False Positive' and `True Positive' alerts and the number of vulnerabilities found by each IAST tool. We also present the vulnerabilities with their respective severity and type based on OWASP Top Ten categories. Additionally, we have compared all our IAST results with those of SAST, DAST, and manual penetration testing from Elder et al.~\cite{Elder2022}.

\subsubsection{Number of Alerts and Vulnerabilities}\label{sec:num-alerts-vuln}
The number of vulnerabilities for each tool was determined by the number of `True Positive' AND `Unique' alerts generated by that tool using the process described in Section \ref{sec:procedure}. We present these results in Table~\ref{tab:vcounts}.  The final row of Table~\ref{tab:vcounts} is the number of vulnerabilities detected by that technique or tool exclusively and not detected by other techniques and tools. The table compares our IAST results with SMPT, EMPT, DAST, and SAST results from the previous work by Elder et al.~\cite{Elder2022}. In the previous work, True Positive alerts were referred to as True Positive \textit{failures}, which applied to both alerts from DAST and SAST as well as failing test cases from SMPT. Elder et al. did not have a true positive count for EMPT comparable to the failing test cases from SMPT or true positive alerts from DAST and SAST. The number of true positive alerts and subsequent ratio for EMPT is marked Not Applicable (N/A). 

\newcommand*{\pColwidth}{0.085\textwidth}
\newcommand*{\iColwidth}{0.082\textwidth}
\newcommand*{\rColwidth}{0.1\textwidth}
\begin{table}[]
\setlength\tabcolsep{3pt}
\centering
\renewcommand{\arraystretch}{1.2}
\begin{tabular}{|B{0.24\textwidth}||T{\pColwidth}|T{\iColwidth} | T{\iColwidth}|| C{\pColwidth}|C{\pColwidth}|C{\pColwidth}|C{\pColwidth}|}
\cline{2-8}
 \multicolumn{1}{c|}{} & \multicolumn{3}{c||}{\textbf{Current Study}} & \multicolumn{4}{c|}{Previous Study}\\\cline{2-8}
\multicolumn{1}{c|}{} & IAST Total & IAST 1 & IAST 2 & SMPT Total & EMPT Total &  DAST Total & SAST Total\\\hline
\hline
True Positive (TP) Alerts & 322 & 159 & 182 & 60 & N/A & 787 & 948 \\\hline
Unique{\newline}Vuln. & 91 & 23 & 68 & 37 & 185 & 23 & 823\\\hline
\hline
Vuln. NOT found by others & 52 & 13 & 41 & 10 & 153 & 5 & 812\\\hline
\end{tabular}
\vspace{1pt}
\caption{Vulnerability Counts (\textit{Section~\ref{sec:num-alerts-vuln}})}
\label{tab:vcounts}
\end{table}

The results in Table~\ref{tab:vcounts} show that IAST finds more unique vulnerabilities than SMPT and DAST, but less than EMPT and SAST. We see a similar trend in the number of vulnerabilities only found by a particular technique or tool. IAST seems to find more vulnerabilities that were not found by other tools compared to SMPT and DAST, but less than EMPT and SAST. We also observe a difference in the number of vulnerabilities found and vulnerabilities unique to each tool for IAST-1 and IAST-2.

\subsubsection{Vulnerability Detection Tool Precision}\label{sec:results-precision}
Precision is calculated as the ratio of the number of `True Positive' alerts to the total number of alerts generated by the tool. As specified in Section~\ref{sec:procedure}, two researchers classified each alert produced by the tools as `True Positive' or `False Positive'.~Any disagreements regarding classification were resolved by a third reviewer. The resultant set consisted of `Total Alerts', `True Positive Alerts', and `False Positive Alerts' generated by each tool. These results are shown in Table ~\ref{tab:tpfp}. The precision of IAST-1 is 0.5, and IAST-2 is 0.34. The IAST results are compared with the automated vulnerability detection tools in Elder et al. ~\cite{Elder2022}. The combined precision of IAST tools of 0.4 is higher than that of DAST tools (0.23) but lower than that of SAST tools (0.98).

\begin{table}[htb!]
\centering
\renewcommand{\arraystretch}{1.2}
\setlength\tabcolsep{2pt}
\begin{tabular}{|B{0.25\textwidth}||T{\pColwidth}|T{\iColwidth} | T{\iColwidth}|| C{\pColwidth}|C{\pColwidth}|}
\cline{2-6}
\multicolumn{1}{c|}{} & \multicolumn{3}{C{0.265\textwidth}||}{\textbf{Current Study}} & \multicolumn{2}{C{0.2\columnwidth}|}{Previous Study}\\\cline{2-6}
\multicolumn{1}{c|}{}& IAST Total & IAST 1 & IAST 2 & DAST Total & SAST Total \\\hline
 \hline
 Total Alerts & 857 & 321 & 536 & 3412 & 962\\\hline
 \hline
 True Positives & 322 & 159 & 182 & 787 & 948\\\hline
 False Positives & 516 & 162 & 354 & 2625 & 20\\\hline
 \hline
 Precision & 0.4 & 0.5 & 0.34 & 0.23 & 0.98 \\\hline
\end{tabular}
\vspace{1pt}
\caption{Total Alerts, False Positives (FP), and Precision for the Current Study Compared with Elder et al. (\textit{Section~\ref{sec:results-precision}})}
\label{tab:tpfp}
\end{table}

\subsubsection{Vulnerability Severity}\label{sec:results-severity}
\begin{figure}[htb!]
\centering
\begin{tikzpicture}[scale=0.8, transform shape]
\begin{scope}
\begin{axis}[
legend cell align={left},
legend style={
  fill opacity=0.8,
  draw opacity=1,
  text opacity=1,
  at={(0.97,0.03)},
  anchor=south east,
  draw=lightgray204
},
tick align=outside,
tick pos=left,
x grid style={darkgray176},
xlabel={\textbf{Number of Vulnerabilities}},
xmin=0, xmax=32.55,
xtick style={color=black},
y dir=reverse,
y grid style={darkgray176},
ylabel={\textbf{OWASP Top Ten Category}},
ymin=-0.5, ymax=9.5,
ytick style={color=black},
ytick={0,1,2,3,4,5,6,7,8,9},
yticklabels={
  01: Broken Access Control,
  02: Cryptographic Failures,
  03: Injection,
  04: Insecure Design,
  05: Security Misconfiguration,
  06: Vulnerable \& Outdated Components,
  07: Identification \& Authentication Failures,
  08: Software \& Data Integrity Failures,
  09: Security Logging \& Monitoring Failures,
  10: Server Side Request Forgery
}
]
\draw[draw=none,fill=red15300] (axis cs:0,-0.25) rectangle (axis cs:14,0.25);
\addlegendimage{xbar,xbar legend,draw=none,fill=red15300}
\addlegendentry{More Severe}

\draw[draw=none,fill=red15300] (axis cs:0,0.75) rectangle (axis cs:4,1.25);
\draw[draw=none,fill=red15300] (axis cs:0,1.75) rectangle (axis cs:11,2.25);
\draw[draw=none,fill=red15300] (axis cs:0,2.75) rectangle (axis cs:9,3.25);
\draw[draw=none,fill=red15300] (axis cs:0,3.75) rectangle (axis cs:2,4.25);
\draw[draw=none,fill=red15300] (axis cs:0,4.75) rectangle (axis cs:0,5.25);
\draw[draw=none,fill=red15300] (axis cs:0,5.75) rectangle (axis cs:11,6.25);
\draw[draw=none,fill=red15300] (axis cs:0,6.75) rectangle (axis cs:5,7.25);
\draw[draw=none,fill=red15300] (axis cs:0,7.75) rectangle (axis cs:3,8.25);
\draw[draw=none,fill=red15300] (axis cs:0,8.75) rectangle (axis cs:0,9.25);
\draw[draw=none,fill=lightblue1] (axis cs:0,-0.25) rectangle (axis cs:0,0.25);
\addlegendimage{xbar,xbar legend,draw=none,fill=lightblue1}
\addlegendentry{Less Severe}

\draw[draw=none,fill=lightblue1] (axis cs:4,0.75) rectangle (axis cs:18,1.25);
\draw[draw=none,fill=lightblue1] (axis cs:11,1.75) rectangle (axis cs:19,2.25);
\draw[draw=none,fill=lightblue1] (axis cs:9,2.75) rectangle (axis cs:26,3.25);
\draw[draw=none,fill=lightblue1] (axis cs:2,3.75) rectangle (axis cs:31,4.25);
\draw[draw=none,fill=lightblue1] (axis cs:0,4.75) rectangle (axis cs:0,5.25);
\draw[draw=none,fill=lightblue1] (axis cs:11,5.75) rectangle (axis cs:12,6.25);
\draw[draw=none,fill=lightblue1] (axis cs:5,6.75) rectangle (axis cs:6,7.25);
\draw[draw=none,fill=lightblue1] (axis cs:3,7.75) rectangle (axis cs:12,8.25);
\draw[draw=none,fill=lightblue1] (axis cs:0,8.75) rectangle (axis cs:0,9.25);
\draw (axis cs:7,0) ++(0pt,0pt) node[
  scale=1,
  text=white,
  rotate=0.0
]{14};
\draw (axis cs:2,1) ++(0pt,0pt) node[
  scale=1,
  text=white,
  rotate=0.0
]{4};
\draw (axis cs:5.5,2) ++(0pt,0pt) node[
  scale=1,
  text=white,
  rotate=0.0
]{11};
\draw (axis cs:4.5,3) ++(0pt,0pt) node[
  scale=1,
  text=white,
  rotate=0.0
]{9};
\draw (axis cs:1,4) ++(0pt,0pt) node[
  scale=1,
  text=white,
  rotate=0.0
]{2};
\draw (axis cs:0.5,5) ++(0pt,0pt) node[
  scale=1,
  text=darkred2,
  rotate=0.0
]{0};
\draw (axis cs:5.5,6) ++(0pt,0pt) node[
  scale=1,
  text=white,
  rotate=0.0
]{11};
\draw (axis cs:2.5,7) ++(0pt,0pt) node[
  scale=1,
  text=white,
  rotate=0.0
]{5};
\draw (axis cs:1.5,8) ++(0pt,0pt) node[
  scale=1,
  text=white,
  rotate=0.0
]{3};
\draw (axis cs:0.5,9) ++(0pt,0pt) node[
  scale=1,
  text=darkred2,
  rotate=0.0
]{0};
\draw (axis cs:14.5,0) ++(0pt,0pt) node[
  scale=1,
  text=cornflowerblue114139207,
  rotate=0.0
]{0};
\draw (axis cs:11,1) ++(0pt,0pt) node[
  scale=1,
  text=black,
  rotate=0.0
]{14};
\draw (axis cs:15,2) ++(0pt,0pt) node[
  scale=1,
  text=black,
  rotate=0.0
]{8};
\draw (axis cs:17.5,3) ++(0pt,0pt) node[
  scale=1,
  text=black,
  rotate=0.0
]{17};
\draw (axis cs:16.5,4) ++(0pt,0pt) node[
  scale=1,
  text=black,
  rotate=0.0
]{29};
\draw (axis cs:1.5,5) ++(0pt,0pt) node[
  scale=1,
  text=cornflowerblue114139207,
  rotate=0.0
]{0};
\draw (axis cs:11.5,6) ++(0pt,0pt) node[
  scale=1,
  text=black,
  rotate=0.0
]{1};
\draw (axis cs:5.5,7) ++(0pt,0pt) node[
  scale=1,
  text=black,
  rotate=0.0
]{1};
\draw (axis cs:7.5,8) ++(0pt,0pt) node[
  scale=1,
  text=black,
  rotate=0.0
]{9};
\draw (axis cs:1.5,9) ++(0pt,0pt) node[
  scale=1,
  text=cornflowerblue114139207,
  rotate=0.0
]{0};
\end{axis}
\end{scope}
\end{tikzpicture}
\caption{Severity of IAST Vulnerabilities grouped into 2021 OWASP Top Ten Categories}
\label{fig:severity_type}
\end{figure}
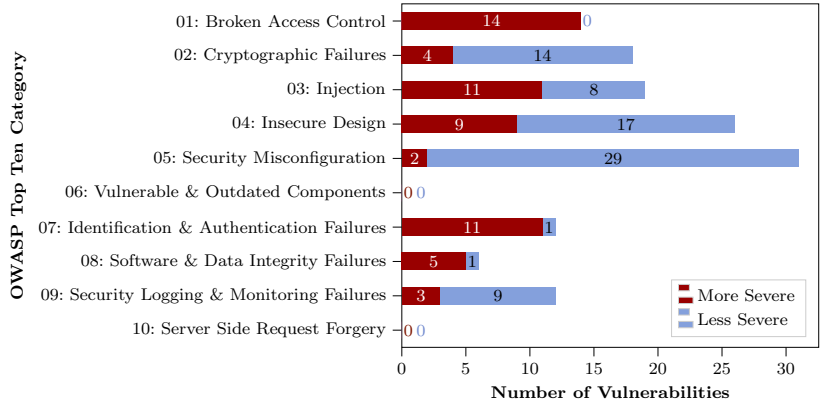

As explained in Section ~\ref{sec:vseverity}, the severity of the vulnerabilities detected by the IAST tools was grouped into `more severe' and `less severe' categories based on the initial severity values assigned to them by the tools. In Figure ~\ref{fig:severity_type}, we  show the number of `more severe' and `less severe' vulnerabilities detected by IAST tools grouped into the OWASP Top Ten (2021) categories as the OWASP Top Ten orders the vulnerabilities types based on how critical they are. We can see from the figure that overall, the number of `less severe' vulnerabilities detected by IAST is greater than the number of 'more severe' vulnerabilities detected. 

We compare the severity of vulnerabilities found by IAST tools with those by Elder et al.~\cite{Elder2022} in Table ~\ref{tab:severity}. The vulnerabilities are distributed over OWASP Top Ten (2021) categories so as to present the vulnerabilities in the most critical ranked fashion. In the case where a vulnerability was assigned more than one OWASP Top Ten category by an IAST tool, the vulnerability was counted towards all the OWASP Top Ten categories assigned to it. 
The comparison of the severity of vulnerabilities, amongst IAST-1 and IAST-2 and overall, has been shown in Table~\ref{tab:severity_iast}. IAST-2 reports a higher number of both `more severe' and `less severe' vulnerabilities than IAST-1.  

\begin{table}[]
\centering
\renewcommand{\arraystretch}{1.2}
\begin{tabular}{|l|c|c|c|}
\hline
Severity    & Overall & IAST-1 & IAST-2 \\ \hline\hline
More Severe & 36 & 13 & 24 \\\hline
Less Severe & 55 & 7 & 51 \\\hline
\end{tabular}
\vspace{1pt}
\caption{IAST Vulnerability Severity (\textit{Section~\ref{sec:results-severity}})}
\label{tab:severity_iast}
\end{table}


\newcommand*{\prevcolwidth}{0.1\textwidth}
\newcommand*{\footwidthi}{0.9\textwidth}
\begin{table}[ht!]
\centering
\renewcommand{\arraystretch}{1.2}
\setlength\tabcolsep{1pt}
\newcounter{prevfoot2}
\setcounter{prevfoot2}{2}
\newcounter{prevfoot3}
\setcounter{prevfoot3}{3}
\newcounter{prevfoot4}
\setcounter{prevfoot4}{4} 
\begin{tabular}{|B{0.35\textwidth}||T{0.1\textwidth}| | C{\prevcolwidth}|C{\prevcolwidth}|C{\prevcolwidth}|C{\prevcolwidth}|}
\cline{2-6}
 \multicolumn{1}{c|}{} & {\textbf{Curr. Study}} & \multicolumn{4}{c|}{Previous Study}\\ \cline{2-6}
 \multicolumn{1}{c|}{}  &  IAST & SMPT & EMPT & DAST & SAST \\\hline
\hline
01: Broken Access Control & 14$^{\mathrm{ab}}$ & 2$^{\dag}$ & 15 & 0 & 28\\\hline
02: Cryptographic Failures & 4 & 1 & 1 & 1 & 2 \\\hline
03: Injection & 11$^{\mathrm{cde}}$ & 5 & 119 & 11 & 24 \\\hline
04: Insecure Design & 9$^{\mathrm{ac}}$ & 5$^{\dag}$ & 8 & 1 & 8 \\\hline
05: Security Misconfiguration & 2$^{\mathrm{d}}$ & 2 & 2 & 2 & 14\\\hline
06: Vulnerable \& Outdated Components & 0 & 0 & 0 & 0 & 0 \\\hline
07: Identification \& Authentication Failures & 11$^{\mathrm{b}}$ & 13 & 10 & 1 & 2 \\\hline
08: Software \& Data Integrity Failures & 5$^{\mathrm{e}}$ & 1$^{\ddag}$ & 0 & 0 & 10\\\hline
09: Security Logging \& Monitoring Failures & 3$^{\mathrm{a}}$  & 3 & 9 & 0 & 0\\\hline
10: Server-side Request Forgery & 0 & 1$^{\ddag}$ & 0 & 0 & 0\\\hline
\hline
No Mapping to OWASP Top Ten & 0 & 1 & 1 & 0 & 54\\\hline
\hline
More Severe Total& 36 & 32 & 165 & 17 & 142\\\hline
\multicolumn{6}{p{\footwidthi}}{\tiny{$^{\mathrm{a}}$Three (3) ``More Severe'' vulnerabilities found using IAST were associated with all three of: A01, A04, and A09}} \\
\multicolumn{6}{p{\footwidthi}}{\tiny{$^{\mathrm{b}}$Eleven (11) ``More Severe'' vulnerabilities found using IAST were associated with both A01 and A07}} \\
\multicolumn{6}{p{\footwidthi}}{\tiny{$^{\mathrm{c}}$Two (2) ``More Severe'' vulnerabilities found using IAST were associated with both A03 and A04}} \\
\multicolumn{6}{p{\footwidthi}}{\tiny{$^{\mathrm{d}}$Two (2) ``More Severe'' vulnerabilities found using IAST were associated with both A03 and A05}} \\
\multicolumn{6}{p{\footwidthi}}{\tiny{$^{\mathrm{e}}$Two (2) ``More Severe'' vulnerabilities found using IAST were associated with both A03 and A08}} \\
\hline
\multicolumn{6}{p{\footwidthi}}{\tiny{\textit{From the Prior Work\cite{Elder2022}:}}} \\
\multicolumn{6}{p{\footwidthi}}{$^{\dag}$\tiny{One vulnerability found using SMPT mapped to both A01 and A04 through two different CWEs.}} \\
\multicolumn{6}{p{\footwidthi}}{$^{\ddag}$\tiny{One vulnerability found using SMPT mapped to both A08 and A10 through two different CWEs.}} \\
\end{tabular}
\vspace{1pt}
\caption{Number of `More Severe' Vulnerabilities Detected by Each Technique based on 2021 OWASP Top Ten (\textit{Section~\ref{sec:results-severity}})}
\label{tab:severity}
\end{table}


\subsubsection{Vulnerability Type (OWASP Top Ten and CWE)}\label{sec:results-vuln-type}
A vulnerability type comparison Elder et al.~\cite{Elder2022} is shown in Table ~\ref{tab:vtype}. The first column of Table ~\ref{tab:vtype} indicates the OWASP Top Ten (2021) categories. The total vulnerability count calculated for each technique or tool is provided in the final `Total'  row. The columns indicate the vulnerability counts for each of the vulnerability detection techniques and tools, distributed over the OWASP Top Ten categories. From the table, we can observe that IAST is most effective in finding A05:Security Misconfiguration vulnerabilities, and the same is evident from Figure~\ref{fig:severity_type}. We also note that RASP is able to prevent only A03:Injection vulnerabilities.

As shown in Table ~\ref{tab:vtype}, none of the vulnerability detection techniques or tools found vulnerabilities in the OWASP Top Ten Category for Vulnerable and Outdated Components (A06), suggesting that different techniques and categories of techniques are useful for finding different types of vulnerabilities. Finding vulnerabilities in components or third-party code of the application requires tools such as Software Composition Analysis (SCA) tools. We have not included these tools in our study and instead focus on the vulnerabilities that arise from the source code of the application itself.

\newcommand*{\typecolwidth}{0.087\textwidth}
\begin{table}[!htb]
\centering
\renewcommand{\arraystretch}{1.2}
\setlength\tabcolsep{1.5pt}
\begin{tabular}{|B{0.35\textwidth}||T{0.094\textwidth}||T{0.094\textwidth}| | C{\typecolwidth}|C{\typecolwidth}|C{\typecolwidth}|C{\typecolwidth}|}
\cline{2-7}
 \multicolumn{1}{c|}{} &  \multicolumn{2}{c||}{\textbf{Curr. Study}} & \multicolumn{4}{c|}{Previous Study}\\\cline{2-7} 
 \multicolumn{1}{c|}{}  &  IAST & RASP & SMPT & EMPT & DAST & SAST \\\hline
\hline
01: Broken Access Control & 14$^{\mathrm{ab}}$ & 0& 2$^{\dag}$ & 15 & 1 & 261\\\hline
02: Cryptographic Failures & 18$^{\mathrm{f}}$ & 0 & 1 & 1 & 2 & 4 \\\hline
03: Injection & 19$^{\mathrm{cdeg}}$ & 44 & 5$^{\dag}$ & 119 & 11 & 58 \\\hline
04: Insecure Design & 26$^{\mathrm{ach}}$ & 0 & 8 & 26 & 2 & 36 \\\hline
05: Security Misconfiguration & 31$^{\mathrm{dfh}}$ & 0 & 5 & 4 & 6 & 15 \\\hline
06: Vulnerable \& Outdated Components & 0  & 0 & 0 & 0 & 0 & 0 \\\hline
07: Identification \& Authentication Failures & 12$^{\mathrm{b}}$ & 0 & 13 & 10 & 1 & 2 \\\hline
08: Software \& Data Integrity Failures & 6$^{\mathrm{e}}$ & 0 & 1$^{\ddag}$ & 0 & 1 & 11\\\hline
09: Security Logging \& Monitoring Failures & 12$^{\mathrm{ag}}$ & 0 & 3 & 9 & 0 & 0\\\hline
10: Server-side Request Forgery & 0 & 0 & 1$^{\ddag}$ & 0 & 0 & 0\\\hline
\hline
No Mapping to OWASP Top Ten & 0 & 0 & 1 & 2 & 0 & 436\\\hline
\hline
Total& 91 & 44 & 37 & 187 & 44 & 823\\\hline
\multicolumn{7}{p{0.9\textwidth}}{\tiny{$^{\mathrm{a}}$Three (3) vulnerabilities found using IAST were associated with all three of: A01, A04, and A09}} \\
\multicolumn{7}{p{0.9\textwidth}}{\tiny{$^{\mathrm{b}}$Eleven (11) vulnerabilities found using IAST were associated with both A01 and A07}} \\
\multicolumn{7}{p{0.9\textwidth}}{\tiny{$^{\mathrm{c}}$Two (2) vulnerabilities found using IAST were associated with both A03 and A04}} \\
\multicolumn{7}{p{0.9\textwidth}}{\tiny{$^{\mathrm{d}}$Two (2) vulnerabilities found using IAST were associated with both A03 and A05}} \\
\multicolumn{7}{p{0.9\textwidth}}{\tiny{$^{\mathrm{e}}$Two (2) vulnerabilities found using IAST were associated with both A03 and A08}} \\
\multicolumn{7}{p{0.9\textwidth}}{\tiny{$^{\mathrm{f}}$Thirteen (13) vulnerabilities found using IAST were associated with both A02 and A05}} \\
\multicolumn{7}{p{0.9\textwidth}}{\tiny{$^{\mathrm{g}}$Eight (8) vulnerabilities found using IAST were associated with both A03 and A09}} \\
\multicolumn{7}{p{0.9\textwidth}}{\tiny{$^{\mathrm{h}}$Three (3) vulnerabilities found using IAST were associated with both A04 and A05}} \\
\hline
\multicolumn{7}{p{0.9\textwidth}}{\tiny{\textit{From the Prior Work\cite{Elder2022}:}}} \\
\multicolumn{7}{p{0.9\textwidth}}{\tiny{$^{\dag}$One vulnerability found using SMPT mapped to both A01 and A04 through two different CWEs.}} \\
\multicolumn{7}{p{0.9\textwidth}}{\tiny{$^{\ddag}$One vulnerability found using SMPT mapped to both A08 and A10 through two different CWEs.}} \\
\end{tabular}
\caption{Vulnerability Type Comparison with Elder et al.~\cite{Elder2022}, including  More Severe \textit{and} Less Severe Vulnerabilities (\textit{Section~\ref{sec:results-vuln-type}})}
\label{tab:vtype}
\end{table}

\bigskip
\begin{tcolorbox}[colback=white,colframe=black!50,title=RQ1 - \rqIASTeffective]
{\color{black!50}\textbf{Answer:}} \textit{IAST finds more unique vulnerabilities than SMPT and DAST, but less than EMPT and SAST. However, IAST found new vulnerabilities that were not found by the techniques.}
\end{tcolorbox}

\subsection{RQ2 : Efficiency of IAST}\label{sec:results-rq2}
We define the efficiency of techniques and tools in terms of vulnerabilities detected per hour (VpH) and calculate it as the ratio of the number of vulnerabilities detected by a tool to the total time taken to classify all the alerts of that tool as `true positive'/`false positive' and `unique'/`duplicate'. We use the average as the statistical measure for obtaining a central value of efficiency for each tool. We present the calculated efficiency of IAST tools in Table ~\ref{tab:efficiency} and compare the results with those of Elder et al.~\cite{Elder2022}.
As per the table, the efficiency of IAST tools of 2.14 VpH is found to be second only to EMPT (2.22 VpH), and even that difference is quite minor. However, the efficiency of the two IAST tools differ significantly, suggesting that the efficiency is highly dependent on the tool.

\newcommand*{\efficwideCol}{0.09\textwidth}
\newcommand*{\efficshortCol}{0.09\textwidth}
\begin{table}[htb!]
\centering
\renewcommand{\arraystretch}{1.2}
\setlength\tabcolsep{1pt}
\begin{tabular}{|B{0.17\textwidth}||T{0.12\textwidth}|T{\efficwideCol} | T{\efficwideCol}|| C{\efficshortCol}|C{\efficshortCol}|C{\efficshortCol}|C{\efficshortCol}|}
\cline{2-8}
\multicolumn{1}{c|}{} & \multicolumn{3}{c||}{\textbf{Current Study}} & \multicolumn{4}{C{0.35\textwidth}|}{Previous Study}\\\cline{2-8}
\multicolumn{1}{c|}{} &  IAST Overall & IAST 1 & IAST 2 & SMPT & EMPT & DAST & SAST \\\hline
\hline
Avg. VpH & 2.14 & 1.34 & 2.98 & 0.69 & 2.22 & 0.55 & 1.17\\\hline
\end{tabular}
\vspace{1pt}
\caption{Efficiency across techniques and tools using Vulnerabilities per Hour (VpH) (\textit{Section~\ref{sec:results-rq2}})}
\label{tab:efficiency}
\end{table}

\bigskip
\begin{tcolorbox}[colback=white,colframe=black!50,title=RQ2 - \rqIASTefficiency  ]
{\color{black!50}\textbf{Answer:}} \textit{The efficiency of IAST tools of 2.14 VpH is found to be second only to EMPT (2.22 VpH), and even that difference is quite minor. However, the efficiency of the two IAST tools also differs, suggesting that the efficiency is highly dependent on the tool. }
\end{tcolorbox}

\subsection{RQ3 : Effectiveness of RASP}\label{sec:results-rq3}
As detailed in Section ~\ref{sec:raspMeth}, we determine the effectiveness of the RASP tool based on two factors: (1) the number of vulnerabilities prevented from being exploited which has been detected by IAST, SMPT, EMPT, DAST, and SAST; (2) the type of vulnerabilities prevented from being exploited by the tool.  We present the types of vulnerabilities prevented by the RASP tool in Table ~\ref{tab:vtype}. RASP prevented, on average, 44 vulnerabilities from being exploited. RASP detects and prevents A03:Injection attacks quite effectively. However, RASP was not able to detect or prevent any attacks caused by the exploitation of any other type of vulnerability. 

\bigskip
\begin{tcolorbox}[colback=white,colframe=black!50,title=RQ3 - \rqRASP  ]
\textbf{\color{black!50}{Answer:}} \textit{RASP prevented, on average, 44 vulnerabilities from being exploited for each execution of the test suite.}
\end{tcolorbox}

\section{Discussion}
\label{chap:disc}
In this section, we discuss our findings and include suggestions on how practitioners might use them to choose among different software vulnerability detection and prevention techniques and tools to make a more informed choice. 
Table~\ref{tab:discussion} gives an overview of the factors discussed in the following sections for selecting software vulnerability detection techniques and tools.

\newcommand*{\techColwidthB}{0.45\textwidth}
\newcommand*{\curColwidthB}{0.45\textwidth}
\begin{table}[ht!]
\centering
\renewcommand{\arraystretch}{1.2}
\setlength\tabcolsep{1pt}

\begin{tabular}{|C{0.14\textwidth}||C{0.125\textwidth}|C{0.18\textwidth}  || C{0.13\textwidth}|C{0.12\textwidth}|C{0.12\textwidth}|C{0.12\textwidth}|}
\cline{2-7}
\cline{2-7}
 \multicolumn{1}{c|}{} & \multicolumn{2}{c||}{Features} & \multicolumn{4}{c|}{Performance (Results)}\\\hline
Technique/{\newline}{Tool} &  {Exploit-ability} & Vuln. Location & Efficiency (VpH) & Unique Vuln. & OWASP Top Ten Types & Precision \\\hline
\hline
IAST & Yes & \textbf{Provided} & 2.14 & 52 & 8 & 0.4\\\hline \hline
SAST & Unknown & \textbf{Provided} & 1.17 & 812 & 7 & 0.98\\\hline
DAST & Yes & Not Provided & 0.55 & 5 & 7 & 0.23 \\\hline
SMPT & Yes & Not Provided & 0.69 & 10 & 9 & N/A\\\hline
EMPT & Yes & Not Provided & 2.22 & 153 & 7 & N/A\\\hline
\end{tabular}
\caption{Summary of Vulnerability Detection Tools (\textit{Section~\ref{chap:disc}})}
\label{tab:discussion}
\end{table}

\subsection{Individual Vulnerability Types (Effectiveness)}
Practitioners having prior knowledge of the prevalence of types of vulnerabilities or wishing to prioritize a particular type of vulnerability in their application may be more motivated to use techniques and tools that are more effective in detecting and preventing that specific type of vulnerability. 
Table~\ref{tab:severity} and Table~\ref{tab:vtype} illustrate how some techniques are more effective in addressing specific types of vulnerabilities. For example, EMPT is most effective in its ability to identify Injection vulnerabilities. We can also see in Table~\ref{tab:vtype} that the RASP tool is the most effective in preventing the exploitation of Injection vulnerabilities present in a running system. Therefore, practitioners focusing on Injection vulnerabilities could use EMPT in the testing phase to detect Injection vulnerabilities in the application and use the RASP tool when deploying their application to block Injection attacks in deployment. 

\subsection{Coverage (Effectiveness)}
For a wide variety of vulnerability types, practitioners may want to emphasize maximum detection coverage. In each of the eight OWASP Top Ten 2021 categories, IAST tools found at least six vulnerabilities, as shown in Table ~\ref{tab:vtype}. Compared to other automated detection techniques, SAST detected five types, and DAST detected two types of at least six vulnerabilities. When compared with manual techniques 
SMPT is found to be better than IAST, which provided the most coverage of the OWASP Top Ten Categories amongst all the techniques and tools, as shown in Table ~\ref{tab:vtype}. Despite the fact that the identical 131 test cases were conducted for both SMPT from Elder et al. ~\cite{Elder2022} and IAST in the current study, each IAST tool found a higher number of ``unique'' vulnerabilities than SMPT as shown in Table~\ref{tab:vcounts}. These findings imply that our study benefited from the use of IAST when conducting black-box test cases. 

Although the RASP tool does not block any other types of vulnerability exploitation apart from A03:Injection using our set of test cases, preventing injection exploits is beneficial. Moreover, the types of attacks prevented by the RASP tool might differ based on the application and the method of attack being used to exploit vulnerabilities in the application.

 Therefore, IAST can effectively detect the number of critical (OWASP Top Ten) vulnerability types second best to only SMPT but still detect more unique vulnerabilities than SMPT. Thus, practitioners prioritizing maximum coverage might want a vulnerability detection technique or tool similar to IAST. Furthermore, practitioners might prefer to employ a RASP tool to deploy additional protection measures in a "Defense in Depth" security strategy.


\subsection{Severity (Effectiveness)}

Our results from Tables ~\ref{tab:severity}, ~\ref{tab:severity_iast} and Figure ~\ref{fig:severity_type} indicate that, unlike the previous section, IAST might not be a good choice when practitioners are aiming for detecting more severe vulnerabilities as the total number of 'more severe' vulnerabilities detected by IAST is only better than DAST. Although EMPT detects the most `more severe' vulnerabilities, these vulnerabilities are primarily Injection vulnerabilities. SAST tool might prove to be a better choice to detect a higher number of `more severe' vulnerabilities across OWASP Top Ten types. Therefore, practitioners prioritizing finding the most number of `more severe' vulnerabilities might want to employ EMPT or SAST. 

\subsection{Automation (Efficiency)}

From the results in Table~\ref{tab:efficiency}, we can see that the efficiency of IAST is second to only EMPT and is better than the other automated tools, SAST and DAST. Studies such as~\cite{Rahman2015} indicate that practitioners might prefer using automated tools for integrating testing into the continuous deployment pipeline for automation and an assumption that manual techniques are less efficient than automated techniques. However, our results indicate that the efficiency of manual techniques is comparable to that of automated ones. 

Thus, practitioners prioritizing efficient resource utilization might want to consider all the techniques, that is, manual as well as automated, to select the most efficient technique or tool based on resource availability.

\section{Threats to Validity and Limitations}
\label{sec:lims}
We discuss the Threats to Validity and Limitations of our work.  We organize this discussion based on four categories of potential threats to Validity: Conclusion Validity, External Validity, Internal Validity, and Construct Validity~\cite{cook1979quasi,feldt2010validity,wohlin2012experimentation}.

\subsection{Conclusion Validity}
Conclusion Validity is about whether conclusions are based on statistical evidence~\cite{cook1979quasi,wohlin2012experimentation}. While we have empirical results, a single case study is insufficient to draw \textit{statistically significant} conclusions for effectiveness and efficiency. Measuring effectiveness with statistical significance would require the application of all four techniques to at least 10-20 additional applications~\cite{kirk2013experimentaldesign}. Applying all techniques to 10-20 similarly-sized SUTs is impractical, given the effort required to apply these techniques to a single application. To mitigate this threat to validity, we performed an extensive review of the vulnerability counts, using the guidelines in Section~\ref{sec:count}, and at least two individuals were involved in the review process for each technique to verify the accuracy of the results. We also assess efficiency for at least two individuals to minimize the risk of introducing bias.

\subsection{Construct Validity}
Construct Validity concerns the extent to which the treatments and outcome measures used in the study reflect the higher level constructs we wish to examine~\cite{cook1979quasi,wohlin2012experimentation,ralph2018construct}. The number of vulnerabilities and types of vulnerabilities are commonly used measures of (in)security in academia and industry \cite{Klees2018,SATEV}, including by the U.S. National Institute of Standards and Technology (NIST) Software Assurance Metrics and Tool Evaluation (SAMATE) program~\cite{SATEV}. We selected our metrics based on these standards and the standards used in the prior work by Elder et al.~\cite{Elder2022} to enable us to compare the performance of the new techniques against well-known approaches.

\subsection{Internal Validity}\label{sec:InternalValidity}
Internal Validity concerns whether the observed outcomes are due to the treatment applied, and whether other factors may have influenced the outcome~\cite{cook1979quasi,feldt2010validity}.
Our primary threat to internal validity is the classification of alerts generated by the tools as `true positive'/`false positive' and `unique'/`duplicate' has been done manually and, therefore, prone to human errors and biases, such as domain knowledge and differences in efficiency of humans. To mitigate this, we have used two reviewers for the classification of alerts and a third reviewer in case of conflicts.

\subsection{External Validity}\label{sec:limitations-external}
External Validity concerns the generalizability of our results~\cite{cook1979quasi,feldt2010validity,wohlin2012experimentation}. One threat to external validity is the selection of tools used in this study may not be generalizable. With the exception of RASP, we have used two tools per technique to accommodate tool differences. However, the results may not represent those by other tools of the same technique. 

Another threat to external validity is the generalizability of our results to other systems. We have used the tools on one large system, OpenMRS, which uses mostly Java and web technologies.  We selected a system that uses a diverse set of technologies, Java, Hibernate, a wide range of web technologies for both frontend and backend, and SQL for databases.  However, our results may not generalize to other systems using other technologies.

\section{Conclusion}
\label{chap:conc}


This study has been motivated by the questions from the practitioners regarding the selection of vulnerability detection and prevention techniques and tools for maximizing the security of applications with optimal resource utilization. The previous study by Elder et al.~\cite{Elder2022} had a similar motivation. However, with the recent introduction and increasing discussion on using IAST and RASP tools, our study was motivated to extend the comparison of four vulnerability detection techniques to include the IAST vulnerability detection tool, as well as the RASP exploit prevention tool. We extend their study by testing these two categories of tools and comparing each vulnerability from our result set with their set of vulnerabilities. The main finding of Elder et al.~\cite{Elder2022} still holds - each approach to vulnerability detection found vulnerabilities NOT found by the other techniques. However, IAST adds value to the range of vulnerability detection techniques available for organizations to use in terms of increasing efficiency and effectiveness in detecting OWASP Top Ten vulnerabilities.

\section{Future Work}
\label{sec:future}

We used a number of metrics, such as vulnerability count, to compare the vulnerability detection and prevention tools on a large application. Based on our experience and study, we found a detailed protocol is required to maintain consistency in the procedure adopted so as to make the results replicable and less subjective. However, more research is required to evaluate what other factors should be considered in a comparison of vulnerability detection techniques. The scope of research is even more relevant for vulnerability prevention tools, which are quite nascent as compared to the earlier vulnerability detection techniques, especially when they are still being improved upon to be integrated into an actual production environment and make the process smoother.  

An additional area of future work is a further exploration of vulnerability severity and related measures, such as the exploitability of vulnerabilities. As can be seen from Table ~\ref{tab:vtype}, the interpretation of severity can vary between measures of criticality. IAST tools themselves detected vulnerabilities that they classified as `low severity' but which were associated with `Broken Access Control', the \#1 most critical vulnerability according to the OWASP Top Ten. Similarly, vulnerabilities associated with disclosure of information about application functionality through error messages, associated with \#4 in the OWASP Top Ten - `Insecure Design' were not considered particularly critical in this context given that the application is open-source~\cite{Elder2022}. Therefore, to understand the severity and exploitability of a vulnerability, more research is required.

\section*{Acknowledgment}

This work was supported by the National Science Foundation [blinded] grant.  The authors would also like to thank [blinded] for their valuable input on this paper.

 \bibliographystyle{elsarticle-num} 
\bibliography{IASTRASP}

\end{document}